\documentclass[aps,prb,twocolumn,superscriptaddress,showpacs]{revtex4}
\usepackage[centertags]{amsmath}
\usepackage{graphicx,amssymb,nicefrac}

\newcommand{\cn}{$\kappa$-(BEDT-TTF)$_2$Cu$_2$(CN)$_3$}
\newcommand{\cnp}{$\kappa'$-(BEDT-TTF)$_2$Cu$_2$(CN)$_3$}
\newcommand{\anion}{$\big[{\rm Cu}_2({\rm CN})_3\big]^-$}
\newcommand{\cy}{CN$^-$}

\begin{document}
\title{Temperature dependence of structural and electronic properties of the spin-liquid candidate $\kappa$-(BEDT-TTF)$_2$Cu$_2$(CN)$_3$}

\author{Harald O. Jeschke}

\affiliation{Institut f\"ur Theoretische Physik, Goethe-Universit\"at Frankfurt am Main, 60438 Frankfurt am Main, Germany}

\author{Mariano de Souza\cite{presentaddress}}

\affiliation{Physikalisches Institut, Goethe-Uni\-ver\-si\-t\"at Frank\-furt am Main, 60438 Frankfurt am Main, Germany}

\author{Roser Valent\'\i}

\affiliation{Institut f\"ur Theoretische Physik, Goethe-Universit\"at Frankfurt am Main, 60438 Frankfurt am Main, Germany}

\author{Rudra Sekhar Manna}

\affiliation{Physikalisches Institut, Goethe-Uni\-ver\-si\-t\"at Frank\-furt am Main, 60438 Frankfurt am Main, Germany}

\author{Michael Lang}

\affiliation{Physikalisches Institut, Goethe-Uni\-ver\-si\-t\"at Frank\-furt am Main, 60438 Frankfurt am Main, Germany}

\author{John A. Schlueter}

\affiliation{Materials Science Division, Argonne National Laboratory, Argonne, Illinois 60439, United States}

\begin{abstract}
  We investigate the effect that the temperature dependence of the
  crystal structure of a two dimensional organic charge-transfer salt
  has on the low-energy Hamiltonian representation of the electronic
  structure. For that, we determine the crystal structure of {\cn} for
  a series of temperatures between $T=5$~K and 300~K by single crystal
  X-ray diffraction and analyze the evolution of the electronic
  structure with temperature by using density functional theory and
  tight binding methods.  We find a considerable temperature
  dependence of the corresponding triangular lattice Hubbard
  Hamiltonian parameters. We conclude that even in the absence of 
  change of symmetry, the temperature dependence of quantities like
  frustration and interaction strength can be significant and should
  be taken into account.
\end{abstract}

\date{\today}

\pacs{
71.15.Mb, % 	Density functional theory, local density approximation, gradient and other corrections
75.10.Jm, %     Quantized spin models, including quantum spin frustration
61.05.cp, %	X-ray diffraction
61.66.Hq  %	Organic compounds
}

\maketitle

\section{Introduction}
The two-dimensional organic charge-transfer salts based on
bis(ethylenedithio)tetrathiafulvalene (BEDT-TTF or even shorter ET)
molecules in a $\kappa$-type lattice arrangement have been intensively
studied over the past thirty years due to their complex interplay
between electron correlation and the effects of low dimensionality and
spin frustration.~\cite{Kanoda11} In particular, the discovery of
spin-liquid behavior in {\cn}~\cite{Shimizu03} has fascinated
experimentalists and theorists alike. Issues of current interest
concern the nature of the low-temperature spin-liquid realized in this
material~\cite{SYamashita08,MYamashita08,Pratt11} and the various
anomalies observed upon approaching the spin-liquid state from high
temperatures. These anomalies include drastic changes in the $^1$H-NMR
relaxation rate around 200-150~K~\cite{Kurosaki05}, the thermopower at
150~K~\cite{Komatsu96}, relaxor-type
%These anomalies include drastic changes in the thermopower~\cite{Komatsu96} and the $^{1}$H-NMR relaxation rate~\cite{Kurosaki05} around 150\,K, relaxor-type 
ferroelectricity
around 60\,K~\cite{Abdel10} and a mysterious phase-transition anomaly
at 6\,K~\cite{Manna10}. The latter feature, which manifests itself in anomalies in
thermodynamic~\cite{SYamashita08,Shimizu03} and
transport~\cite{MYamashita08} quantities, is accompanied by pronounced
lattice effects.~\cite{Manna10} Various scenarios have been suggested
for the 6\,K anomaly including a crossover from a thermally to a
quantum disordered state~\cite{SYamashita08}, an instability of the
quantum spin-liquid~\cite{SYamashita08, Baskaran89, Liu05, Kyung06,
  Grover10, Lee07, Galitski07, Kawamura84} or a distinct type of
charge ordering.~\cite{Li10} Theoretically, the spin-liquid
properties have been investigated on the basis of the anisotropic
triangular-lattice Hubbard
Hamiltonian.~\cite{Kyung06,Mizusaki06,Tocchio09,Yang10,Balents10}  The
parameters $t$, $t'$ and $U$ of this Hamiltonian have been determined
with semi-empirical~\cite{Komatsu96} as well as first principles
methods~\cite{Kandpal09,Nakamura09} based on the experimental
structure at room temperature.  Missing in such investigations is,
however, the consideration of a possible temperature dependence of the
model parameters.

%The temperature dependence of the
%electronic properties, in particular the phase diagram, has so far
%been studied based on a temperature independent
%Hamiltonian~\cite{Kyung06,Mizusaki06}.
In this work, we find by a combination of single crystal X-ray
diffraction at various temperatures and density functional theory
calculations that even in the absence of structural phase transitions,
the temperature dependence of the structural parameters is significant
enough to influence the electronic behavior and the determination of
the Hamiltonian model parameters in {\cn}. We suggest that this has
subtle effects on the degree of frustration and interaction strength.

%warrant determination of Hubbard Hamiltonian parameters at various
%temperatures based on the temperature evolution of the crystal
%structures.

In 1991, the crystal structure of {\cn} was first
reported,~\cite{Geiser91} and the unit cell parameters
confirmed,~\cite{Komatsu91} on twinned crystals at room
temperature. The crystal structure was redetermined in 1993, also on a
twinned crystal at room temperature.~\cite{Papavassiliou93} The
crystal structure of the very similar {\cnp} structure, which is
reported to be an ambient pressure superconductor, was described in
1992.~\cite{Yamochi92,Yamochi93} The crystal structure of the $\kappa$ phase was
subsequently redetermined in 1997,~\cite{Bu97} and the unit cell
redetermined in 2001, both at room temperature.~\cite{Drozdova01} Already in
1991, the unit cell parameters were reported at 300 and 30~K, with
$a_{(30\,K)}/a_{(300\,K)} =0.9964$, $b_{(30\,K)}/b_{(300\,K)} =
0.9932$, and $c_{(30\,K)}/c_{(300\,K)} = 0.9900$, and
$\beta_{(30\,K)}/\beta_{(300\,K)} = 1.0146$.~\cite{Bu93} 
%While this report suggests that the $c$-axis has the greatest relative contraction with temperature, we find that the $a$-axis has the greatest relative contraction over the 300 to 20~K temperature range with
Our comparable contraction values between 300 and 20~K are
$a_{(20\,K)}/a_{(300\,K)} =0.9988$, $b_{(20\,K)}/b_{(300\,K)} =
0.9958$, and $c_{(20\,K)}/c_{(300\,K)} = 0.9905$, and
$\beta_{(20\,K)}/\beta_{(300\,K)} = 1.0148$ and agree well with the
previous results. In addition, we find evidence for an ordering of the
ethylene groups in a staggered conformation between 200-150~K. Despite
the current interest in this material as a spin-liquid candidate, no
low-temperature structural determinations have yet been
reported. Herein, we present a detailed characterization of the
crystal structure as function of temperature determined on a single
crystal with agreement factors better than any previously reported.

\section{Single crystal X-ray diffraction}
A black, plate-like crystal of {\cn} with dimension $0.4 \times 0.4
\times 0.4$~mm$^3$ was placed onto the tip of a glass fiber and
mounted on a Bruker APEX II 3-circle diffractometer equipped with an
APEX II detector. Temperature control in the 100-300\,K region was
provided by an Oxford Cryostream 700 Plus Cooler while below 100\,K it
was provided by a Cryocool-LHE cryogenic system (Cryo Industries of
America). The sample temperature below 100 K was confirmed by installing a Cernox thermometer (Lakeshore) in the immediate vicinity of the crystal and stabilizing the temperature immediately prior to data collection.
%The sample temperature was accurately monitored employing a Cernox thermometer (Lakeshore) installed in the immediate vicinity of the sample. 
In order to reduce adverse thermal effects the thermometer wires
(twisted manganin wires, 0.5~mm diameter, supplied by Lakeshore) were
anchored on the surfaces of the cryostat exposed to $^4$He stream.
The data were collected 
%with 0.3$^\circ$ omega scans 
using MoK$_\alpha$ radiation ($\lambda = 0.71073$~{\AA}) with a
detector distance of 50~mm. Unit cell parameters were determined upon
cooling in 10~K increments between 100 and 290\,K, with a frame
exposure time of 10 seconds. Full data sets for structural analysis
were collected at temperatures of 5, 20, 100, 150, 200, 250 and
300\,K. The uncertainties in the temperature determination are
typically $\pm0.2$~K down to 100~K, while for the 20~K and 5~K data
points an error bar of $\pm 1$~K has to be accepted.  These data
collections nominally covered over a hemisphere of reciprocal space by
a combination of three sets of exposures.
%; each set had a different phi angle for the crystal. 
Data to a resolution of 0.68~{\AA} were considered in the
reduction. The raw intensity data were corrected for absorption
(SADABS~\cite{Sheldrick01a}). The structure was solved and refined
using SHELXTL.~\cite{Sheldrick01b} A direct-method solution was
calculated, which provided most of atomic positions from the
E-map. Full-matrix least squares / difference Fourier cycles were
performed, which located the remaining atoms. All non-hydrogen atoms
were refined with anisotropic displacement parameters. The hydrogen
atoms were placed in ideal positions and refined as riding atoms with
relative isotropic displacement parameters.

\begin{table*}[htb]
\begin{tabular}{rrrrrrrr}
\hline
                    &{\bf 300 K} &{\bf 250 K}  &{\bf 200 K}  &{\bf 150 K} &{\bf 100 K} &{\bf 20 K} &{\bf 5 K}\\\hline\hline
$a$ ({\AA})         &16.0919(3)  &16.0848(3)   &16.0781(3)   &16.0703(3)  &16.0746(6)  &16.072(4)  &16.062(3)\\
$b$ ({\AA})         &8.5722(2)   &8.5749(1)    &8.5737(1)    &8.5664(2)   &8.5593(3)   &8.536(2)   &8.544(2)\\
$c$ ({\AA})         &13.3889(2)  &13.3373(2)   &13.2964(2)   &13.2698(3)  &13.2678(5)  &13.262(3)  &13.271(2)\\
$\beta$ ($^\circ$)   &113.406(1)  &113.853(1)   &114.273(1)   &114.609(1)  &14.852(1)   &115.088(3) &115.093(2)\\
$V$ ({\AA}$^3$)     &1694.93(6)  &1682.43(4)    &1670.86(4)  &1660.72(6)  &1656.51(1)   &1647.8(6)  &1649.3(5)\\
$\rho$ (g/cm$^{-3})$ &1.909       &1.923         &1.937       &1.949      &1.954        &1.964      &1.962\\
$\mu$ (mm$^{-1})$    &2.266       &2.283         &2.299       &2.313      &2.319        &2.331      &2.329\\
GoF                 &1.057       &1.053         &1.091       &1.157       &1.293       &1.261       &1.109\\
$R$                 &0.0311      &0.0269        &0.0238      &0.0225      &0.0220      &0.0182      &0.0204\\
$R_w$               &0.0838      &0.0735        &0.0654      &0.0627      &0.0629      &0.0503      &0.0515\\
staggered (\%)      &77          &86            &93          &100         &100         &100         &100\\
$d_{\rm Cu-NNC}$~(\AA)&0.050       &0.056         &0.064       &0.069       &0.073       &0.072       &0.072\\
ET tilt angle $\vartheta$ &66.55 &66.25         &66.00       &65.79       &65.64       &65.53       &65.52\\
$d_{\rm intradimer}$   &3.558      &3.538         &3.518       &3.500       &3.488       &3.470       &3.473
\end{tabular}
\caption{Crystal data and structure refinement of {\cn}.
  Formula = C$_{23}$H$_{16}$Cu$_2$N$_3$S$_{16}$;
  formula weight $M_W = 974.43$, monoclinic, wavelength 
  $\lambda= 0.71073$~{\AA}, effective number of electrons in 
  the crystal unit cell contributing to F(000) = 978,
  space group $P\,2_1/c$, $Z = 2$.
  Residual factor for the reflections $R_1 = \sum \big||F_o|-|F_c|\big|/\sum|F_o|$;
  weighted residual factors 
  $wR_2= \big[\sum w(F_o^2-F_c^2)^2/\sum w(F_o^2)^2\big]^{\nicefrac{1}{2}}$;
  $[I>2\sigma(I)]$;
  least-squares goodness-of-fit parameter 
  ${\rm GoF} = \big[\sum w(F_o^2-F_c^2)^2/(N_d-N_p)\big]^{\nicefrac{1}{2}}$.
  Staggered (\%) is the percentage of ethylene groups that are in the 
  staggered conformation. $d_{\rm Cu-NNC}$~({\AA}) is the distance that 
  the Cu(I) ion is out of the plane defined by the coordinated N, N and C atoms. 
  The ET tilt angle $\vartheta$ is measured against the $bc$ plane. 
  $d_{\rm intradimer}$ is the orthogonal distance between BEDT-TTF dimers.
}\label{tab:lattice}
\end{table*}

\begin{figure}[htb]
\includegraphics[width=0.96\columnwidth]{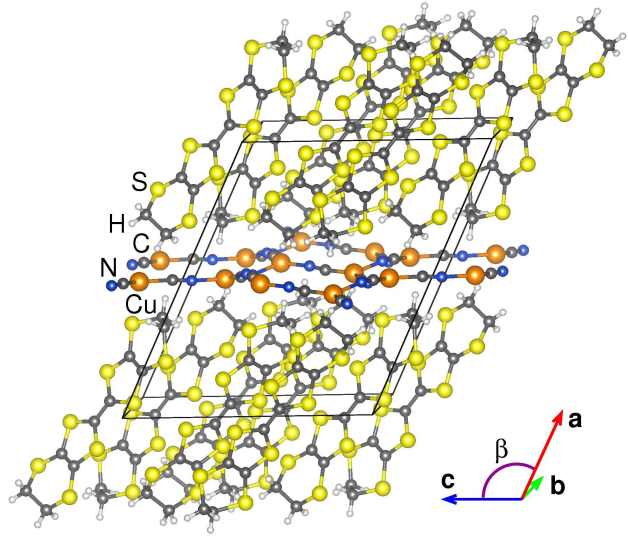}
\caption{Structure of {\cn} at $T=5$~K. Note that the disordered
  {\cy} group in the inversion center is only shown in one
  conformation.}
\label{fig:cryst}
\end{figure}

As a representative example, in Fig.~\ref{fig:cryst} we show the
structure of {\cn} at a temperature of $T=5$~K. In the $bc$ plane, the
BEDT-TTF molecules form their typical $\kappa$-type arrangement. One
of the cyanide ({\cy}) groups resides on an inversion center, thus
requiring it to be disordered with a 50{\%} carbon and 50{\%} nitrogen
distribution on these two atomic positions~\cite{Geiser91}. Analysis
of the low-temperature structural data (see Tab.~\ref{tab:lattice})
indicates that overall, $a$, $b$ and $c$-axes decrease with
temperature while the $\beta$ angle increases monotonically upon
cooling. We find that the orthogonal projection of the $a$-axis,
$a_\perp=a\sin\beta$, has the greatest relative contraction with
temperature.
%Interestingly, the monoclinic angle shows a clear maximum at about 190~K, see Supplementary Information (SI). 
One ethylene group is disordered at room temperature with a staggered
conformation 77{\%} of the time. As the temperature is lowered, the
ethylene group remains partially disordered down to a temperature of 200~K  and is fully ordered  in a staggered conformation at 150~K.
The anion layer becomes slightly more buckled at low temperature: at
room temperature, the Cu1 atom lies 0.050~{\AA} out of the C11-C12-N12
plane, increasing to 0.072~{\AA} at low temperature.  For the purpose
of performing density functional theory calculations, the original
symmetry of $P\,2_1/c$ has been lowered to $P\,2_1$ by choosing one of
two possible orientations of the {\cy} group in the inversion center.

\section{Temperature-dependent structural parameters}
\begin{figure}[htb]
\includegraphics[angle=-90,width=0.96\columnwidth]{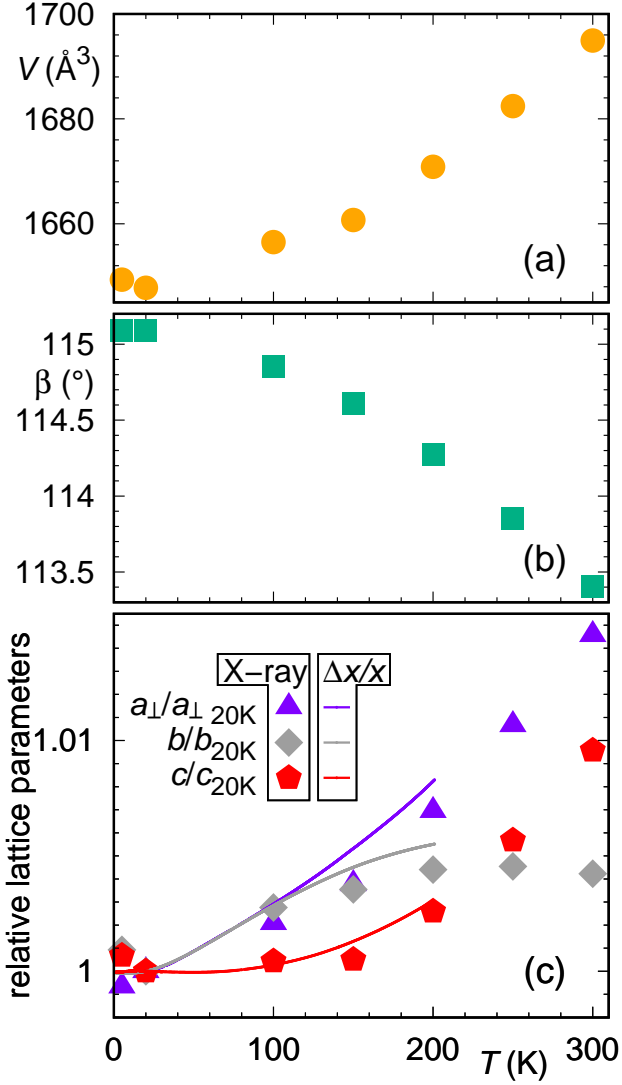}
\caption{Structural parameters of {\cn} between $T=5$~K and
  $T=300$~K. (a) and (b) show the volume and monoclinic angle,
  respectively. In (c) relative lattice parameters are given with the
  $T=20$~K structure as reference. $a_\perp=a\sin\beta$. Symbols refer to the new data from
  X-ray diffraction (this work) while lines are measured thermal
  expansion data from Ref.~\onlinecite{Manna10}. 
%For transforming the $a$-axis data from the measured data along $a^*$, see text.  
  The experimental error bar is comparable to the size of the symbols. }
\label{fig:structure}
\end{figure}

In Fig.~\ref{fig:structure}, we show the evolution of the lattice
parameters with temperature. Over the large investigated temperature
range from $T=300$~K down to $T=5$~K, the volume is monotonously
decreasing with temperature (see Fig.~\ref{fig:structure}~(a)). The
monoclinic angle $\beta$ between ${\bf a}$ and ${\bf c}$ lattice
vectors (see Fig.~\ref{fig:structure}~(b)) first rapidly increases
upon cooling down to a temperature of $T=200$~K, below which it
increases more gradually. In Fig.~\ref{fig:structure}~(c), the lattice
parameters are displayed as symbols, normalized by their values at
$T=20$~K (see Tab.~\ref{tab:lattice}). We also include the relative
$b$ and $c$ lattice constants along the two principal axes as obtained
by thermal expansion measurements in Refs.~\onlinecite{Manna10} and
\onlinecite{deSouza08}. The out-of-plane expansivity data shown there
were taken in a direction perpendicular to the $bc$ plane. They are
show together with the corresponding quantity from the X-ray
diffraction measurement, $a_\perp=a\sin\beta$.

%{\it i.e.} parallel to $a^*$, $\alpha_{a^*}$, were transformed according 
%to the relation $\alpha_{a^*} = \alpha_a\cos^2(\beta-\nicefrac{\pi}{2}) + 
%\alpha_c\sin^2(\beta-\nicefrac{\pi}{2})$, see Ref.~\onlinecite{Barron80}.

%in derived calculated as
%\begin{equation}
%\frac{x(T)}{x(T_0)} = \exp \int_{T_0}^T dT \,\alpha_x(T)
%\end{equation}
%for $T_0=5$~K and $x=a,b,c$ where $\alpha_x(T)$ is the thermal expansion
%\begin{equation}
%\alpha_x(T) = \frac{1}{x(T)}  \frac{dx(T)}{dT}
%\end{equation}
%as measured in Ref.~\onlinecite{Manna10} and \onlinecite{deSouza08}.

\section{Temperature-dependent electronic structure}
\begin{figure}[htb]
\includegraphics[angle=-90,width=0.96\columnwidth]{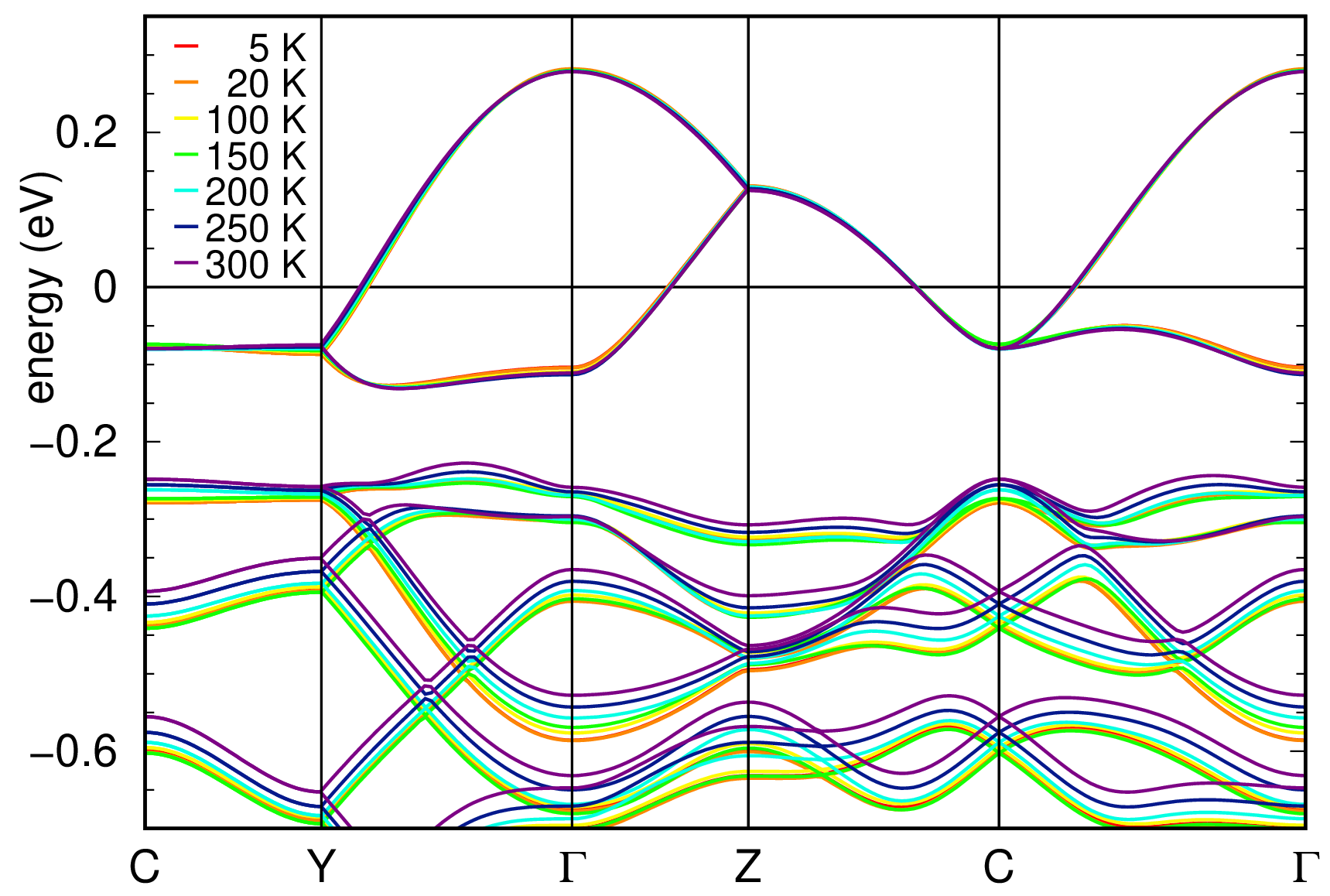}
\caption{Band structures of {\cn} from $T=5$~K to $T=300$~K. }
\label{fig:allbs}
\end{figure}

In the following, we determine the electronic properties for the
resolved crystal structures at different temperatures by employing the
all-electron full-potential local orbitals (FPLO) \cite{Koepernik99}
basis.  We perform all calculations on a $(6\times6\times6)$ $k$ mesh
with a generalized gradient approximation functional.~\cite{Perdew96}
In Fig.~\ref{fig:allbs} we present the electronic band structures for
the various crystal structures. In the calculation, we used the
staggered (majority) conformation of the BEDT-TTF molecules at all
temperatures. The changes as function of temperature for the two bands
at the Fermi level, corresponding to the antibonding combinations of
the BEDT-TTF highest occupied molecular orbital (HOMO) levels, are
relatively small. On the other hand, the occupied bands down to
$-0.7$~eV below the Fermi level show a significant dependence on
temperature.  These bands derive from the bonding combination of
BEDT-TTF HOMO levels and from the {\anion} anion layer. Overall, these
bands show a bandwidth that decreases with increasing
temperature. This can be explained by the volume increase as function
of temperature (see Fig.~\ref{fig:structure}~(a)).

\begin{figure}[htb]
\includegraphics[angle=-90,width=0.96\columnwidth]{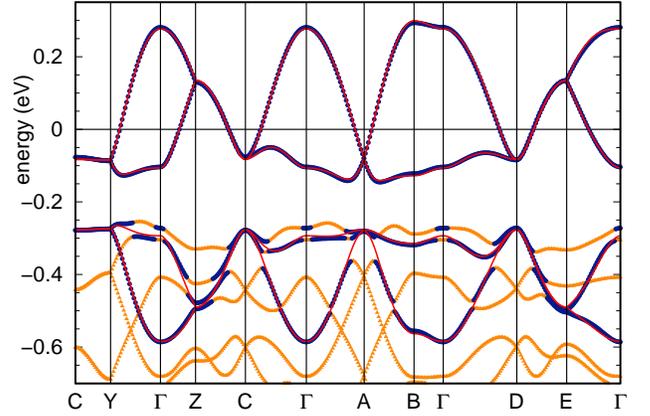}
\caption{Band structure of {\cn} at $T=20$~K. Blue circles (orange
  triangles) indicate bands with a majority of BEDT-TTF and {\anion}
  character, respectively. The TB fit is shown with lines. }
\label{fig:20Kbs}
\end{figure}

\begin{figure}[htb]
\includegraphics[width=0.9\columnwidth]{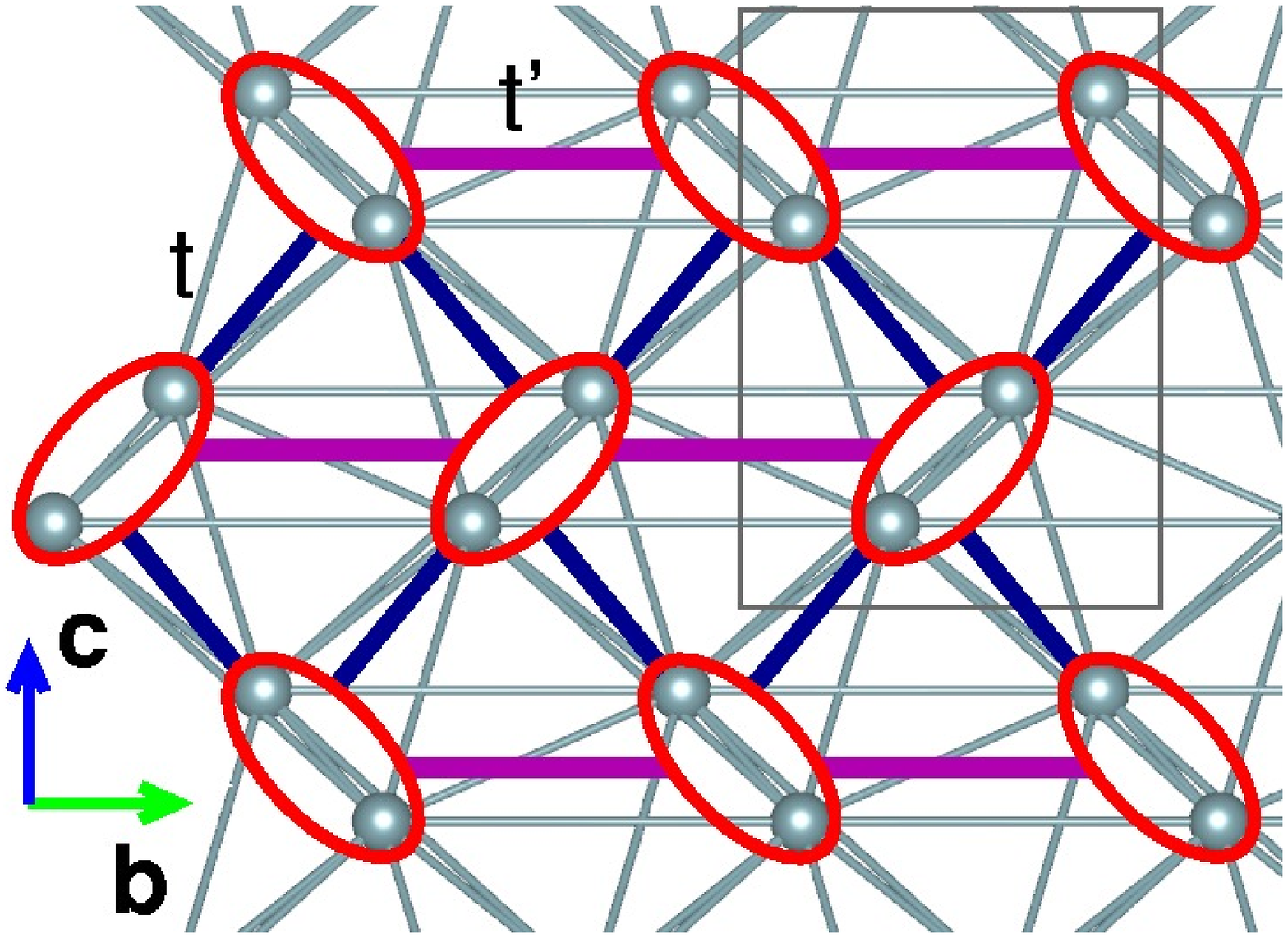}
\caption{Network formed by the BEDT-TTF molecules in the $bc$ plane of
  {\cn}. Each grey ball represents the center of gravity of a BEDT-TTF
  molecule, with grey bonds indicating the intermolecular distances
  that were taken into account for the tight binding fit. BEDT-TTF
  dimers are highlighted by ellipses, and the paths forming the
  triangular lattice paths are shown with bold lines. The unit cell is
  marked as a rectangle.}
\label{fig:network}
\end{figure}

Further analysis of the electronic structure requires a reliable
identification of the bands deriving from the BEDT-TTF molecules. For
that purpose, band weights have been calculated for all structures and
added up for all atoms corresponding to the BEDT-TTF cation layers and
to the {\anion} anion layers, respectively. In Fig.~\ref{fig:20Kbs},
blue circles and orange triangles stand for a predominance of BEDT-TTF
weight and {\anion} weight, respectively. This identification allows
us to fit the BEDT-TTF derived bands to a tight binding Hamiltonian
\begin{equation}
H_{\rm TB} =\sum_{ij,\sigma} t_{i-j} \big(c_{i\sigma}^\dag
c_{j\sigma}^{\phantom{\dag}}+{\rm H.c.}\big)\,,
\end{equation}
where $c_{i\sigma}^\dag$ ($c_{i\sigma}$) create (annihilate) an
electron with spin $\sigma$ at site $i$; the sites correspond to the
positions of the BEDT-TTF molecules, shown as balls in
Fig.~\ref{fig:network}. While a good overall fit of the band structure
can be achieved by including six neighbor BEDT-TTF molecule distances
up to $d=9.4$~{\AA}, a very good fit describing also the small
dispersion along the $B-\Gamma$ direction, as shown by lines in
Fig.~\ref{fig:20Kbs}, requires fourteen neighbor distances up to
$d=14.5$~{\AA}.

We now proceed to analyze the temperature dependence of the tight
binding parameters corresponding to the network of BEDT-TTF dimers
that are highlighted in Fig.~\ref{fig:network}.  These dimers form a
triangular lattice, with the hopping parameter $t'$ connecting dimers
to chains along the $b$ direction and the hopping parameter $t$
forming the $2D$ connections in $c$ direction.  We are interested in
estimating the parameters of the Hubbard Hamiltonian for the
anisotropic triangular lattice
\begin{equation}\begin{split}
    H =& \sum_{<ij>,\sigma} t \big(c_{i\sigma}^\dag
    c_{j\sigma}^{\phantom{\dag}}+{\rm H.c.}\big)+\sum_{[ij],\sigma} t'
    \big(c_{i\sigma}^\dag c_{j\sigma}^{\phantom{\dag}}+{\rm
      H.c.}\big)\\&+U\sum_i \Big(n_{i\uparrow}
    -\frac{1}{2}\Big)\Big(n_{i\downarrow}
    -\frac{1}{2}\Big)\label{eq:Hub}
\end{split}\end{equation}
where $<ij>$ and $[ij]$ indicate summations over nearest and
next-nearest neighbors, respectively.  $t$ and $t'$ can be obtained
from the molecular overlap integrals $t_2$ to $t_4$ by considering the
geometrical formulas
\begin{equation}
t\approx \frac{t_2+t_4}{2}\,,\qquad t'\approx \frac{t_3}{2}.
\end{equation}
For the definition of the overlap integrals $t_n=t_{i-j}$, see
Ref.~\onlinecite{Kandpal09}.  Note that inclusion of longer-range
hopping into these formulas (for example including $t_5$ into $t'$)
has no influence on the results reported in the following.

\section{Discussion}

\begin{figure}[htb]
\includegraphics[angle=-90,width=0.96\columnwidth]{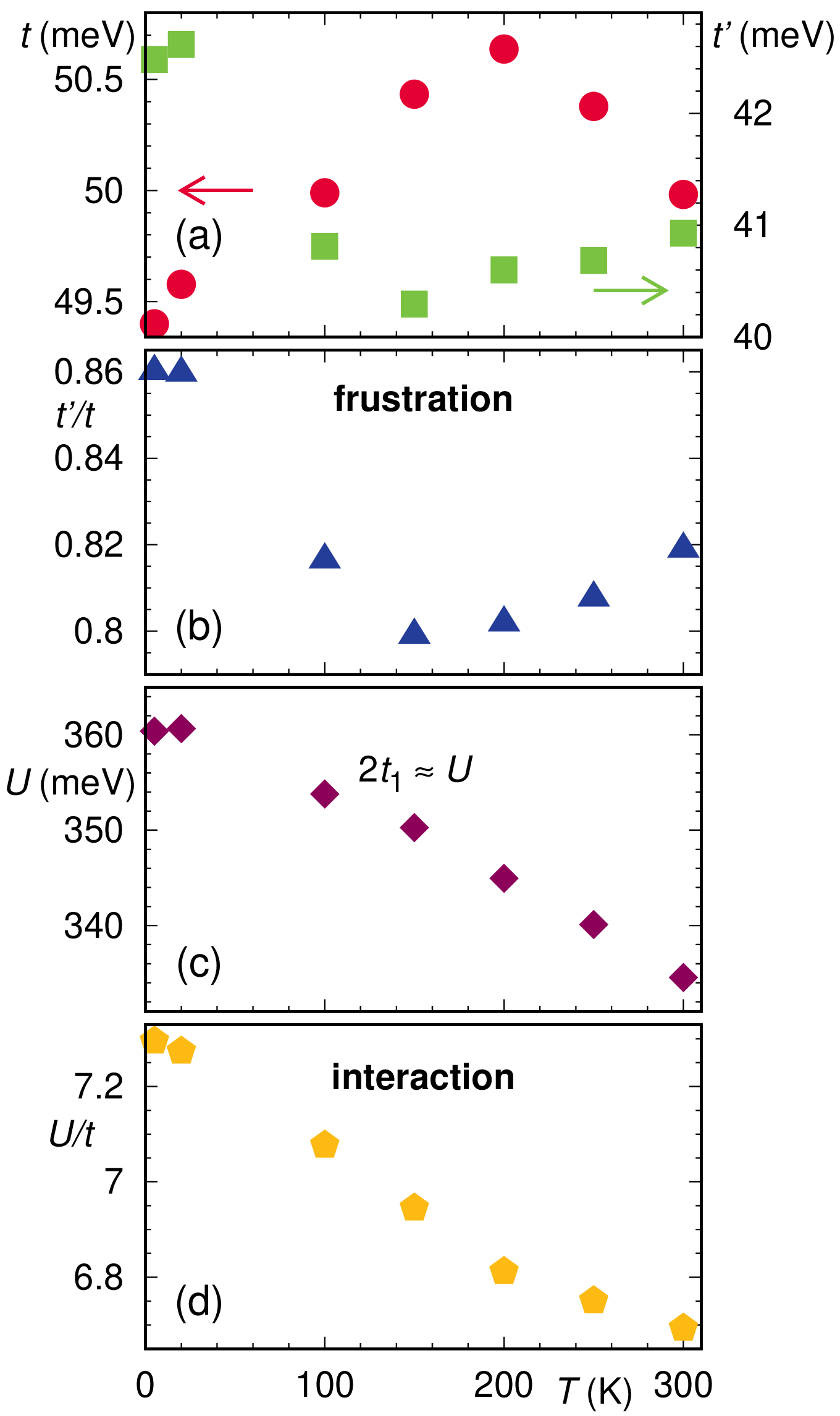}
\caption{Temperature dependence of Hamiltonian parameters of {\cn}. }
\label{fig:parT}
\end{figure}

Fig.~\ref{fig:parT} summarizes our findings from the tight binding
analysis. The nearest-neighbor hopping parameters $t$, forming a
square lattice, increase upon cooling down to a temperature of
$T=200$~K, then decrease again (see Fig.~\ref{fig:parT}~(a)). The
frustrating hopping parameters $t'$ show the opposite behavior as a
function of temperature, decreasing upon cooling down to a temperature of
$T=150$~K, then increasing. Interestingly, these two effects enhance each
other when we consider their ratio $t'/t$ as shown in
Fig.~\ref{fig:parT}~(b)). $t'/t$ which quantifies the degree of
frustration in the system decreases from $t'/t=0.82$ at $T=300$~K to
$t'/t=0.80$ at $T=150$~K, then increases to a maximum value of
$t'/t=0.86$ at $T=5$~K.  A rough estimate for the Coulomb interaction
strength $U$ can also be extracted from the dimer approximation
$U\approx 2t_1$ where $t_1$ is the BEDT-TTF intradimer hopping
integral (see Fig.~\ref{fig:parT}~(c)). We find that the measure of
the interaction strength $U/t$ estimated in this way monotonously
falls by 7\% as the temperature is increased from $T=5$~K to
$T=300$~K.

%The highly nonmonotonous behavior of the overlap integrals $t$, $t'$ and their ratio $t'/t$ is governed by
%should be a sufficiently large effect to have some consequences for
%the behavior of the material as a function of temperature.
%Pronounced features around 150~K-200~K can also be observed in various
%quantities such as spin-lattice relaxation rate divided by
%temperature~\cite{Kurosaki05}, thermoelectric power and spin
%susceptibility~\cite{Komatsu96} as well as a broad bump like anomaly
%in the thermal expansion coefficient along the
%c-axis~\cite{deSouza08}.

\begin{figure}[htb]
\includegraphics[width=0.96\columnwidth]{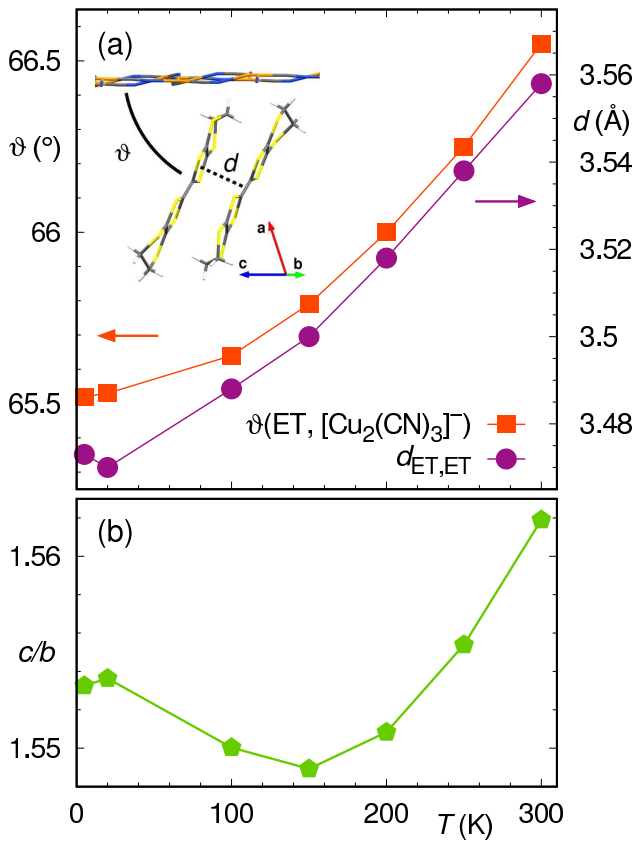}
\caption{(a) Orientation $\vartheta$ and dimerization $d$ of BEDT-TTF
  molecules as function of temperature. The inset illustrates how
  $\vartheta$ and $d$ are measured. (b) Ratio $c/b$ between $c$ and
  $b$ lattice vectors as function of temperature. }
\label{fig:orient}
\end{figure}

In order to rationalize the observed temperature dependence of the
Hamiltonian parameters in \eqref{eq:Hub} we analyze the crystal
structure in more detail. For that purpose, we determined the
orientation of the BEDT-TTF molecules in space by finding the plane of
the TTF part of the molecule and measuring its angle with respect to
the $bc$ plane, cf. inset of Fig.~\ref{fig:orient}. This yields the
inclination of the BEDT-TTF molecules against the anion layer shown as
squares in Fig.~\ref{fig:orient} and the intradimer distance shown as
circles (see also Table~\ref{tab:lattice}). Both quantities show a
nearly monotonous increase over the studied temperature range.  The
decreasing intradimer distance explains the increase of the intradimer
hopping integral $t_1$ with decreasing temperature.  Apparently, the
nonmonotonous evolution of the overlap integrals $t$, $t'$, especially
the distinct extrema in both quantities around 150~K to 200~K, has to
be of different origin.

The overall trend can be understood by considering the temperature
dependence of the lattice parameters; from Fig.~\ref{fig:network} it
is clear that changes in the $b$ lattice parameter should have an
impact on $t'$, while changes in $c$ should affect $t$. In
Fig.~\ref{fig:orient}~(b), we see that the $c/b$ lattice parameter
ratio decreases with temperature down to $T=150$~K, then increases
with falling temperature until $T=20$~K. This has an immediate impact
on the degree of frustration $t'/t$ as it should be approximately
proportional to the $c/b$ ratio. Indeed, comparison of
Fig.~\ref{fig:parT}~(b) and Fig.~\ref{fig:orient}~(b) confirms this
expectation and thus explains the nonmonotonous temperature dependence
of the frustation.

An increasing frustration upon cooling, reaching $t'/t$ values at low
temperatures in excess of those at room temperature, is an interesting
finding which may help to better understand the intriguing
low-temperature magnetic and dielectric properties of this
material. Here we mention the distinct types of charge ordering,
accompanied by dielectric anomalies, proposed in
Ref.~\onlinecite{Li10} for the present material as a result of an
increasing degree of frustration.

The nonmonotonous evolution of the in-plane distortion $c/b$, which
adopts a minimum around 150 K, might be related to the ordering of the
ethylene groups, uncovered in the present study.~\cite{note} According
to our structural analysis, the fraction of ethylene groups, ordered
in the staggered conformation, gradually grows from 77{\%} at room
temperature over 86{\%} at 250~K to 93{\%} at 200~K. For temperatures
below 150 K, the ordering is complete within the accuracy of our
analysis/refinement. The large step size of 50 Kelvin employed
in this study does not allow to determine the ordering temperature
more precisely. Likewise, we cannot say whether or not the ordering
occurs continuously or sets in abruptly. An argument in favour of the
latter possibility might be derived from a small step-like feature
revealed in thermal expansion measurements around 150~K, see the
out-of plane data shown in Fig.~1 of Ref.~\onlinecite{Manna10}. We
stress, that a fully ordered staggered ethylene conformation below
150~K and a progressively disordered state above 200~K, is fully
consistent with the anomalous behavior revealed by $^1$H-NMR
measurements between 150-200~K.~\cite{Kurosaki05} The strong increase
in $(T_1T)^{-1}$ above 200~K was attributed to thermally activated
vibrations of ethylene groups.~\cite{Kurosaki05} We suggest that the
nonmonotonous temperature dependence in $t'/t$ might also be related
to the drastic change in the thermopower around 150
K.~\cite{Komatsu96} The thermopower is related to the energy
derivative of the density of states at the Fermi
level.~\cite{Miyake05} However, for a strongly correlated system like
{\cn}, density functional theory is not sufficient for the calculation
of this quantity and more elaborate many-body calculations -- which
are beyond the scope of the present study -- have to be done.

In summary, we performed an analysis of the temperature dependence of
the structural and electronic properties of {\cn} by considering a
combination of X-ray diffraction at various temperatures and density
functional calculations. Our study shows that the temperature
dependence of the structural parameters has significant influence on
the electronic properties and results in a nonmonotonous behavior of
the degree of frustration. Of special relevance is the increase of
frustration at low temperatures in comparison to the behavior at room
temperature.

\begin{acknowledgments}
  This work was supported by UChicago Argonne, LLC, Operator of
  Argonne National Laboratory (``Argonne''). Argonne, a U.S. Department
  of Energy Office of Science laboratory, is operated under Contract
  No. DE-AC02-06CH11357. We also acknowledge support by the Deutsche
  Forschungsgemeinschaft (SFB/TR~49) and by the Helmholtz Association
  through HA216/EMMI.
\end{acknowledgments}

\appendix
%\section{SUPPLEMENTARY MATERIAL}
\section{Additional crystallographic material}

Crystallographic data for the {\cn} structure at 5, 10, 100, 150, 200,
250 and 300~K has been deposited with the Cambridge Crystallographic
Data Centre as supplementary publication nos. CCDC 850022 to
850028. Copies of the data can be obtained free of charge on
application to CCDC, 12 Union Road, Cambridge CB2 1EZ, UK (fax: (44)
1223 336-033; e-mail: \verb!data_request@ccdc.cam.ac.uk!).

\begin{figure}[htb]
\includegraphics[width=0.96\columnwidth]{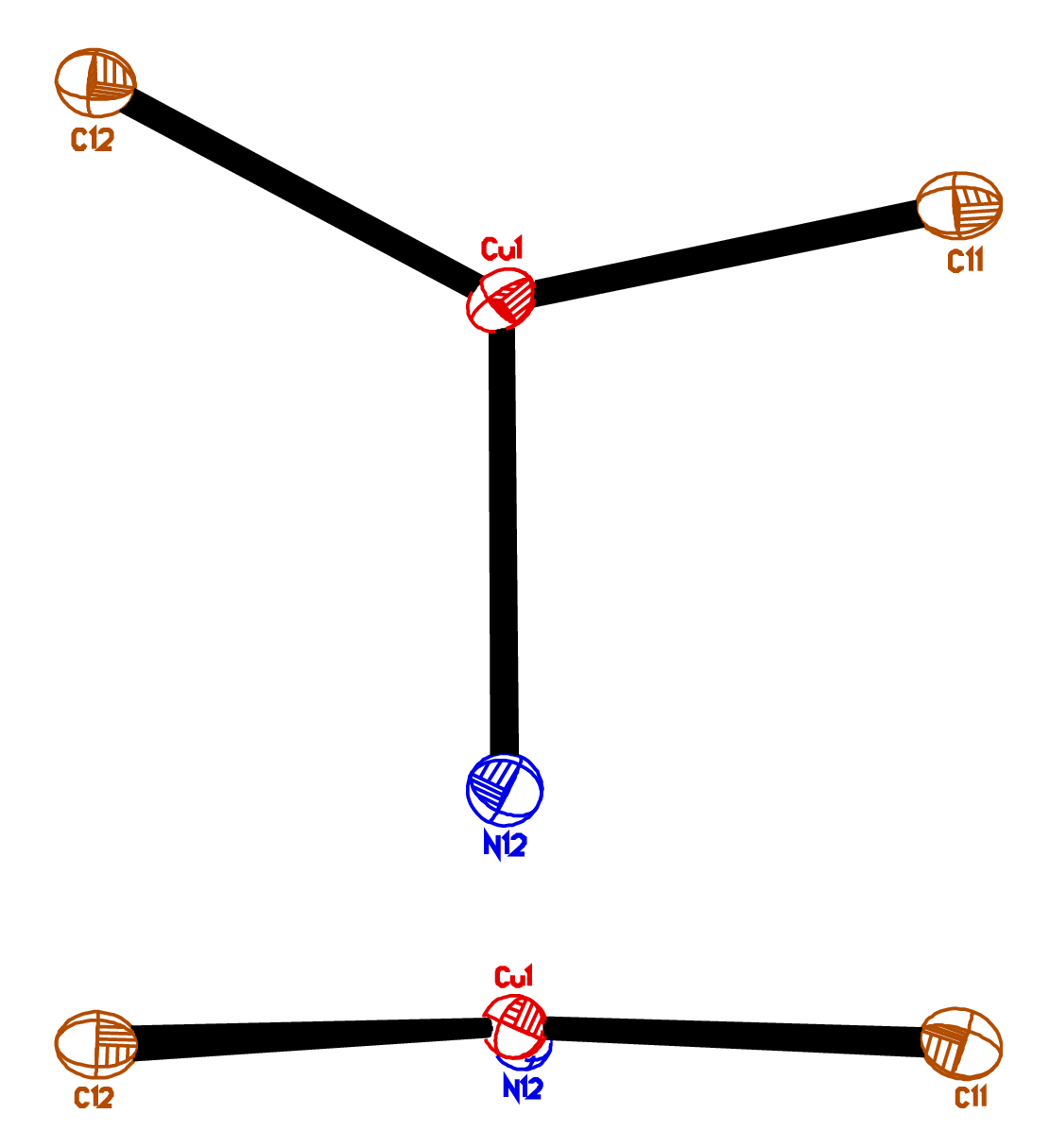}
\caption{The Cu1 coordination sphere at 5~K. The Cu1 is 0.072~{\AA}
  out of the plane defined by N12, C11 and C12. Thermal ellipsoids are
  drawn at the 50\% probability level.}
\label{fig:S1}
\end{figure}

\begin{figure}[htb]
\includegraphics[angle=-90,width=0.96\columnwidth]{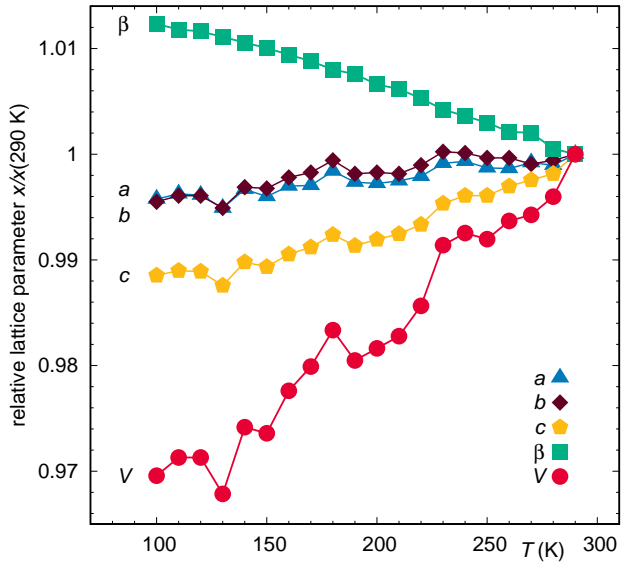}
\caption{Plot of the unit cell parameters as a function of temperature
  between 100 and 290 K.}
\label{fig:S2}
\end{figure}

Fig.~\ref{fig:S1} illustrates geometry of the anion laywer and the
slight deviation from planarity. Fig.~\ref{fig:S2} and
Table~\ref{tab:unitcell} contain lattice parameters determined at 10~K
intervals between 290~K and 100~K.

\begin{table}
\begin{tabular}{cccccc}
T (K)~&~ $a$~(\AA) ~&~ $b$~(\AA) ~&~ $c$~(\AA) ~&~ $\beta$~($^\circ$) ~&~ $V$~(\AA$^3$)\\\hline
290&16.248&8.655&13.516&113.56&1742\\
280&16.231&8.650&13.491&113.62&1735\\
270&16.235&8.647&13.483&113.79&1732\\
260&16.226&8.652&13.475&113.80&1731\\
250&16.227&8.652&13.463&113.90&1728\\
240&16.237&8.656&13.463&113.97&1729\\
230&16.234&8.657&13.453&114.04&1727\\
220&16.214&8.646&13.426&114.16&1717\\
210&16.207&8.639&13.414&114.26&1712\\
200&16.203&8.640&13.407&114.31&1710\\
190&16.205&8.639&13.399&114.42&1708\\
180&16.222&8.650&13.413&114.47&1713\\
170&16.200&8.640&13.397&114.56&1707\\
160&16.199&8.636&13.388&114.63&1703\\
150&16.183&8.627&13.372&114.70&1696\\
140&16.194&8.628&13.378&114.76&1697\\
130&16.165&8.611&13.348&114.82&1686\\
120&16.185&8.621&13.366&114.88&1692\\
110&16.187&8.621&13.367&114.90&1692\\
100&16.179&8.616&13.361&114.96&1689
\end{tabular}
\caption{Unit cell parameters as a function of temperature in the
  temperature range from 290~K down to 100~K.}\label{tab:unitcell}
\end{table}


\begin{thebibliography}{99}
\bibitem{note}
Note that even though all our DFT calculations were performed considering
ordered ethylene groups, they capture the effects of the ethylene ordering
through the changes caused on the structural parameters.
\end{thebibliography}
\end{document}